\documentclass[aps,prd,twocolumn,showpacs]{revtex4}
\usepackage{amsmath}
\usepackage{epsfig}
\usepackage{amsfonts}
\begin{document}
\def\b{\bar}
\def\d{\partial}
\def\D{\Delta}
\def\cD{{\cal D}}
\def\cK{{\cal K}}
\def\f{\varphi}
\def\g{\gamma}
\def\G{\Gamma}
\def\l{\lambda}
\def\L{\Lambda}
\def\M{{\Cal M}}
\def\m{\mu}
\def\n{\nu}
\def\p{\psi}
\def\q{\b q}
\def\r{\rho}
\def\t{\tau}
\def\x{\phi}
\def\X{\~\xi}
\def\~{\widetilde}
\def\h{\eta}
\def\bZ{\bar Z}
\def\cY{\bar Y}
\def\bY3{\bar Y_{,3}}
\def\Y3{Y_{,3}}
\def\z{\zeta}
\def\Z{{\b\zeta}}
\def\Y{{\bar Y}}
\def\cZ{{\bar Z}}
\def\`{\dot}
\def\be{\begin{equation}}
\def\ee{\end{equation}}
\def\bea{\begin{eqnarray}}
\def\eea{\end{eqnarray}}
\def\half{\frac{1}{2}}
\def\fn{\footnote}
\def\bh{black hole \ }
\def\cL{{\cal L}}
\def\cH{{\cal H}}
\def\cF{{\cal F}}
\def\cP{{\cal P}}
\def\cM{{\cal M}}
\def\ik{ik}
\def\mn{{\mu\nu}}
\def\a{\alpha}

\title{New Path to Unification of Gravity with Particle Physics}
\author{Alexander Burinskii}

\affiliation{Laboratory of Theor. Phys. , NSI, Russian Academy of
Sciences, B. Tulskaya 52  Moscow 115191 Russia, \footnote{email:
bur@ibrae.ac.ru}}



\begin{abstract}
    The principal new point is that ultra-high spin of the elementary particles makes Einstein's gravity so strong, that its influence to metric is shifted from Planck to the Compton scale! Compatibility of the Kerr-Newman (KN) gravity with quantum theory is achieved by  implementation of the supersymmetric Higgs model without modification of the Einstein-Maxwell gravity. We consider the nonperturbative bag-like solution to supersymmetric generalized LG field model, which creates a flat and supersymmetric vacuum state inside the bag, forming the  Compton zone for consistent work of quantum theory.  The bag is deformable, and its shape is controlled by BPS bound, providing compatibility of the bag boundary with external gravitational and electromagnetic (EM) field. In particular, for the spinning KN gravity bag
   takes the form of oblate disk with a circular string placed on the disk border. Excitations of the KN EM field create circular traveling waves. The super-bag solution is naturally upgraded to the Wess-Zumino supersymmetric QED model, indicating a bridge from the nonperturbative super-bag  to perturbative formalism of the conventional QED.
\end{abstract}

 \maketitle

\maketitle

\newpage

\bigskip
\section{Introduction}
Modern physics is based on Quantum theory and Gravity. The both theories are confirmed
experimentally with great precision. Nevertheless, they are contradicting and cannot be combined in
a unified theory. One of the principal points is the structure of elementary particles, which are
considered as pointlike and even structureless (for example electron) in  quantum theory, but
should be presented as the extended field configurations for compatibility with the right hide side
of the Einstein equations, $G_\mn = 8\pi T_\mn .$

Revolutionary step for unification was made in superstring theory, however, as mentioned John
Schwarz, \emph{``...Since 1974 superstring theory stopped to be considered as particle physics...
''}  and \emph{``... a realistic model of elementary particles still appears to be a distant dream
... ''}  \cite{Schw}). One of the reasons of this is that extra dimensions are compactified with
extra tiny radii of order the Planck length $10^{-33}$ cm, which does not correlate with
characteristic lengths of quantum physics and  makes impossible to test extra dimensions with
currently available energies. The idea to bring  fundamental gravitational scale close to the weak
scale was considered in different approaches, and in particular, in the brane world scenario, where
the weakness of the localized 4d gravity is explained by its ``leaks'' into the higher-dimensional
bulk, and the brane world mechanism  allowed to realize ideas of the superstring theory for any
numbers of the extra dimensions \cite{DvalBW}.

Alternative ideas were related with nonperturbative 4D solutions of the non-linear field models --
solitons, in particular, solitonic solutions to low energy string theory \cite{Dabh,Sen,BurSen}.
This approach, being akin to the Higgs mechanism of symmetry breaking, is matched with
nonperturbative approach to electroweak sector of the Standard Model. The most known is the
Nielsen-Olesen model of dual string based on the Landau-Ginzburg (LG) field model for a phase
transition in superconducting media, and also the famous MIT and SLAC bag models
\cite{MIT,SLAC,Dash}  which are similar to solitons, but being soft, deformable and oscillating,
acquire many properties of the dual string models. Besides, being suggested for confinement of
quarks, the bag models assume consistent implementation of the Dirac equation. The question on
consistency with gravity is not discussed usually for the solitonic models, as it is conventionally
assumed that gravity is very weak and is not essential on the scale of electroweak interactions.
For example, in \cite{Malda} we read \emph{"... quantum gravity effects are usually very
small, due to the weakness of gravity relative to other forces. Because the effects
of gravity are proportional to the mass, or the energy of the particle, they grow
at high energies. At energies of the order of E ~ $10^{19} $ GeV, gravity would have a
strength comparable with that of the other Standard Model interactions."}

 Our principal point here is that the assumption on the weakness of gravity is not correct, since it is
 based on the underestimation of the role of spin in gravitational interactions.
 Indeed, nobody says that gravity is weak in Cosmology where physics is determined by giant masses.
Similarly, the giant spin/mass ratio of spinning particles makes influence of gravity very strong
in the particle physics.

 For the great spin/mass ratio of the elementary particles, about $ 10^{20}- 10^{22} $   (in dimensionless units $G=c=\hbar =1 $), the commonly accepted view that gravity is weak and not essential in particle physics up to Planck
scale, should be replaced by principally new point of view that \emph{GRAVITY IS NOT WEAK}, and its
influence becomes crucial for the structure of the spinning particles at the Compton scale of the
electroweak interactions.

 We show that spin of the Kerr-Newman (KN) rotating black hole (BH) with parameters of an
electron deforms space-time in the Compton zone so strongly that the compatible with gravity structure of spinning particle
is determined almost unambiguously as a supersymmetric bag model.

The KN space-time  with ultra-high spin has the naked singular ring creating two-sheeted topology. This space differs from Minkowski space so
strongly that neither the Dirac theory nor perturbative QED can be applied, since they require the
flat space at least in the Compton zone of the dressed particle. We show here that  this conflict can be resolved \emph{without modification of the Einstein-Maxwell gravity -- the space can be
cured by a supersymmetric bag model}, in which the singular region of KN solution is replaced by the flat
internal space of the  Compton size. We find the corresponding non-perturbative BPS-saturated
solution in frame of the supersymmetric generalized LG model, in which boundary of the bag is formed
by the domain wall (DW) interpolating between the external KN gravity and the supersymmetric vacuum
state inside the bag. Similar to the usual bag models, the super-bag model is deformable and displays a
super-consistency with the external gravitational and electromagnetic (EM) KN field, in the sense
that its shape and dynamics are fully defined by matching its boundary with a special surface (which can
be called as  "zero gravity surface" (ZG)), where the external gravitational field is compensated by
 the EM field. The ZG-surface determines position of  the  BPS DW-solution, and therefore, it
 determines shape of the bag, as a disk-like configuration with a closed string placed at the sharp border
 of the disk \cite{BurBag,BurBag1,BurPit50El}. We show
that the supersymmetric LG model can be naturally upgraded to the Wess-Zumino SuperQED model \cite{WesBag},
revealing connections between the non-perturbative solutions of the supersymmetric LG model and the
conventional perturbative technics used in QED.

\section{Super-bag model as BPS solution to generalized LG model}

\subsection{Basic features of the ultra-rotating Kerr-Newman solution}
It has been recently obtained \cite{BurSol,BurSol1}, that the source of ultra-spinning
Kerr-Newman (KN) solution can be considered as a superconducting soliton having many features
 of the bag model \cite{BurBag,BurBag1,Bur50}, but
with the essential advantage of compatibility with  Einstein-Maxwell gravity in four dimensions. As
is known, the bag models take intermediate position between strings and solitons
\cite{Giles,JT,Tye}. Although,  the bags  were initially offered  as the extended models of
hadrons, \cite{MIT,SLAC,Dash}, being based on the Abelian  Higgs model of symmetry breaking their
indicated rather applicability to the Salam-Weinberg  model of leptons, which was one of the
reasons to consider the gravitating KN bag as the model for consistent with gravity leptons.

 The spinning KN  solution is of particular interest in this regard, since, as it was obtained
 by Carter \cite{DKS,Car}, that gyromagnetic ratio of the KN solution is $g=2 ,$  and therefore
 corresponds to the external field of the electron. The spin/mass ratio of
 the electron is about $10^{22} ,$ and structure of source of the
 KN solution for such a huge spin  should shed the light on origin of
 the conflict between gravity and quantum theory. One can see that
 the KN field with parameters of electron becomes extremely strong  on
 the Compton distances, so that  the BH horizons disappear and the  Kerr singular
 ring of the Compton radius  $a = \hbar/m $ becomes open, which  breaks topology of space-time
 and creates two-sheeted metric.

 \noindent In the Kerr-Schild (KS) approach, metric of the KN solutions is \cite{DKS}
\be g_\mn =\eta_\mn + 2H k_\m k_\n , \label{KS}\ee where $ \eta_\mn $ is metric of an auxiliary
Minkowski space $M^4 ,$ (signature $(- + + +)$),
and $H$ is the scalar function which for the KN solution takes the form
 \be H_{KN}=\frac {mr - e^2/2}{r^2+a^2 \cos ^2 \theta} ,\label{HKN} \ee where
 $r$ and $\theta$ are oblate spheroidal coordinates, and $ k_\m $ is a null vector
field $ k_\m k^\m =0 ,$ forming a Kerr congruence -- the vortex of polarization of gravitational and electromagnetic field in the Kerr space-time.  The Kerr singular ring corresponds to border of the disk $r=0 , $ in the equatorial plane $\cos\theta =0 .$

Similarly, vector potential of KN solution is also collinear with the null direction $k_\m ,$
  \be A_{\m} = - \frac {e r} { (r^2 + a^2 \cos^2 \theta)} k_\m
\label{Amuk}  \ee

\begin{figure}[ht]
\centerline{\epsfig{figure=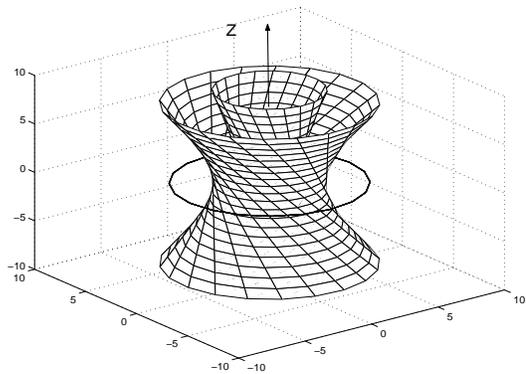,height=5cm,width=7cm}}
\caption{Vortex of the Kerr light-like (null) congruence $k^\m$  propagates analytically from negative sheet of Kerr metric,
$r<0 ,$ to positive one, $r>0 $. In the equatorial plane, $\cos\theta=0 ,$  the Kerr congruence is
focused on the Kerr singular ring, $r=\cos\theta=0 $.}
\end{figure}

The KN metric becomes two-sheeted, since the Kerr congruence \be k_\m dx^\m = dr - dt - a \sin ^2
\theta d\phi,  \label{km} \ee  is out-going at the `positive' sheet of the metric, $r>0 ,$ and
passes analytically to `negative' sheet, $r<0 ,$  being extended via ring $r=0,$  where it becomes
in-going. The two null vector fields $k_\m(x)^\pm$ become different at $r>0$ and $r<0 ,$ leading to
two different metrics $ g_\mn^\pm =\eta_\mn + 2H k_\m^\pm k_\n^\pm $ on the positive and negative
sheet of the same Minkowski background. Similarly, it leads also to two-sheeted vector-potential
$A_{\m}^\pm ,$ that makes space inappropriate for quantum theory, and therefore, conflict between
quantum theory
 and gravity is shifted by 22 orders earlier then it is usually expected, from the Planck to the Compton scale.
 As usually, singularity is signal to new physics --  theory of more high level.
 The KN gravitational field is strong near the Kerr singular ring and creates vortex of the space-time polarization
   in the Compton zone of the dressed electron, which should be flat for work of quantum theory. It is
usually assumed that in vicinity of strong field, gravity  should be modified to a new Quantum Gravity.
Taking into account sharp incompatibility of Quantum and Gravity, natural requirement for such new theory
would be separation of their zones of influence: formation of the internal zone

\textbf{(I)} --
 flat core for quantum theory, and external zone

 \textbf{(E)} -- for undisturbed gravitational and electromagnetic
fields.

There should also be selected intermediate zone

\textbf{(R)} --  interpolating between \textbf{(I)} and \textbf{(E)}.

\noindent In the case of strong KN field, these demands become so restrictive that determine  structure of
the new theory almost uniquely.  It turns out that the flat Compton zone free from gravity may be
achieved \emph{without modification of the Einstein-Maxwell equations,} through SUPERSYMMETRY,
which eats up  the strong gravitational field in the core of particle. Expelling gravity from the
core of the KN spinning particle
 is similar to expelling the EM field from superconducting core, and  both of these super-phenomena are
 realized in core of the KN solution by the supersymmertric LG field model
\cite{FMVW,HLosShifm,AbrTown,CvQRey,CvGrRey,GibTown,ChibShif,WesBag} in the form of a BPS-saturated Super-Bag solution, for which just
the strong contradiction between Quantum and Gravity determines extreme sensitivity of the model to
the choice of the separating surface \textbf{(R)}.

The natural choice of this surface for the KN solution  was suggested by C. L\'opez \cite{Lop}.
According (\ref{KS}) and (\ref{HKN}) it should be the surface "zero gravity" (ZG) at
\begin{equation} r = R = \frac {e^2}{2m} , \label{Hre}
\end{equation}
where function $H $ vanishes \be H_{KN}(R)=0 ,\label{HKNzg} \ee and metric becomes flat, and can be
matched with flat Minkowski space for $r < R .$  It turns the L\'opez source of the KN solution in
a shell-like bubble.
  So far as $r$ is the oblate spheroidal coordinate, \cite{DKS}, related with Cartesian coordinates
  by transformations
  \be x+iy = (r + ia) \exp \{i\phi_K \} \sin \theta , \quad
z=r\cos \theta, \quad \rho =r-t , \label{coordKerr} \ee
  the bubble surface $r=R$ takes the oblate ellipsoidal form -- the disk of the
  thickness $ R $ and radius  $r_c = \sqrt {R^2 + a^2} ,$ where
  $a=J/m .$

For solution without rotation, $a=0 ,$ and bubble turns into a sphere of the classical radius $ r_e .$
Such spherical shape was suggested by Dirac in \cite{DirBag} as an "extensible electron model" -- prototype of the bag
models, displaying one of their basic features of the bags -- their deformability.

We see that deformations of the KN Super-Bag appear as consequence of the requirement on sharp separation of the zones
\textbf{(I),(E),(R)}.

\begin{figure}[ht]
\centerline{\epsfig{figure=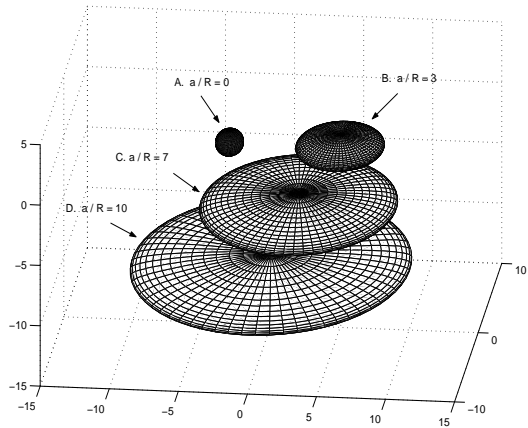,height=6cm,width=7cm}}
\caption{(A):  Spherical bag without rotation $a/R =0$, and  disk-like bags for different values of the rotation parameter: (B)- $ a/R =3$; \ (C) - $ a/R =7$; and (D) - $ a/R =10$.}
\label{fig2}
\end{figure}

\subsection{Spinning bag creates a string}

Usually, it is assumed that bags are deformed by rotations taking the shape of  a string-like  flux-tube joining the
quark-antiquark pair \cite{MIT}.

In the KN Super-Bag, the spinning gravitational field controls disk-like shape of the bag, and
string-like structure is formed  for $a/R >0 $, at  edge rim of the disk, as shown  in Fig.2. In
the equatorial plane, this string approaches  very close to the Kerr singular ring, see Fig.3A, so,
it is really just the singular ring regularized by the bag boundary.

\begin{figure}[ht]
\centerline{\epsfig{figure=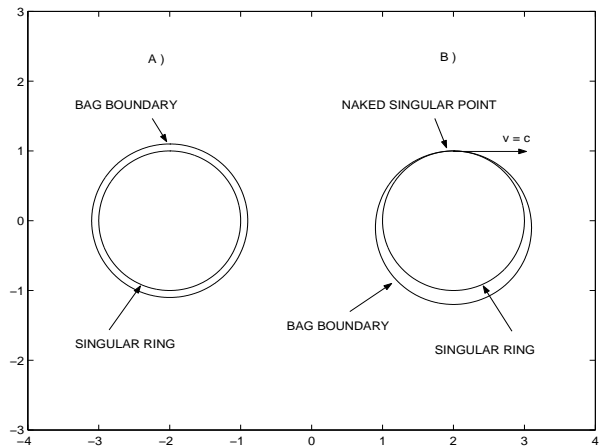,height=6cm,width=8cm}} \caption{Regularization of the KN string.  Boundary of  bag fixes cut-off $R=r_e $ for the Kerr singular ring. A) The exact KN solution. B) The KN solution is excited by the lowest traveling mode: emergence of the singular pole.} \label{fig.7}
\end{figure}

Among diverse attempts to use nonperturbative models in the electroweak sector of the Standard
Model (SM) \cite{Colem,AchVach,Kus,Dval,Volk}, the central place takes the  Nielsen-Olesen (NO) model
\cite{NO,Amb} of the string, which is created as a vortex line in a superconductor.

 The assumption, that Kerr singular ring is similar to NO model of dual string was done very
 long ago in \cite{IvBur,Bur0}, where it was noted that excitations of the KN solution create traveling waves along
 the Kerr ring. Later, it was obtained  in \cite{BurSen,KerSen}  close connection of the Kerr singular ring
 with the Sen fundamental string solution to low energy string theory.\fn{Note also the complex N=2 critical string which was obtained in the complex Kerr geometry \cite{BurStr}.} In the  KN bag model this
 string is formed at the sharp boundary of the superconducting disk, as a dual analog of the NO vortex line in
 superconductor.

 In accordance with the condition (\ref{HKNzg}), the KN gravity  controls position
  of the bag boundary \textbf{(R)}, and also more thin effects, such as excitations of the KN gravity
   define dynamics of the bag and appearance of the traveling waves.

  In particular, it has been shown \cite{BurBag1}, that the lowest EM excitation of the KN solution
  creates the traveling wave which has a circulating lightlike node. At this point, surface of the deformed bag
  touches the Kerr singular ring, as it is shown in Fig.3B, which breaks
regularization at this point and creates the lightlike singular pole, which can be considered as
emergence of the bare Dirac particle circulating inside the Compton zone of dressed electron. On
the other hand, this pole breaks homogeneity of the closed circular string, creating the frontal
and rear ends turning this string in the open. As usual, the end points of an open  string are
associated with quarks, and the KN super-bag model turns into a single ``bag-string-quark'' system,
4D analog of D2-D1-D0-brane system of the string--M-theory.

\section{Supersymmetry  ensures consistency with gravity}

\subsection{Generalized LG field model and  domain wall (DW) phase transition}
 The  LG field model of superconductivity is used in many solitonic models, in particular, in the  NO dual string model, as a field model in the MIT and SLAC bag models, and really, it is  also the the Higgs model of symmetry breaking, because
 the Higgs vacuum itself  "... is analog to a superconducting metal", \cite{Dash}.
  The LG Lagrangian used in the NO model (minimal LG model) is
 \be
{\cal L}_{NO}= -\frac 14 F_\mn F^\mn - \frac 12 (\cD_\m \Phi)(\cD^\m \Phi)^* - V(|\Phi|),
\label{LNO}\ee where $ \cD_\m = \nabla_\m +ie A_\m $  are the $U(1)$ covariant derivatives, and
$F_\mn = A_{\m,\n} - A_{\n,\m} $ is the corresponding field strength, and potential $V$ has  the
quartic form \be V = \lambda ( \Phi ^\dag \Phi - \eta^2)^2 , \label{VNO} \ee
  where $\eta$ is condensate of the Higgs field  $\Phi ,$ its vacuum expectation value (vev) $\eta <|\Phi|> $, \cite{NO}.

The minimal LG model can be used to describe superconductivity inside the bag -- interplay of the KN vector-potential with the Higgs condensate.
 Since requirements \textbf{(I),(E),(R)} define inside the bag a flat space, the corresponding covariant derivatives can be taken as flat,
\be \cD_\m =
\nabla_\m +ie A_\m \ \rightarrow \ \cD_\m = \d_\m +ie A_\m \label{Dflat} . \ee

However, the NO and KN models have opposite spacial configurations: the KN bag model should describe a superconducting disk surrounded by the long-range EM and gravitational field, while the NO model describes vortex of the EM field inside the superconducting Higgs condensate which breaks the external long-range EM and gravitational field.
 Note, that this is a typical drawback of the most of soliton models and, in particular,  the usual bag models which are formed as a "cavity in superconductor" \cite{Dash}. The reason of this disadvantage lies in the use of the
potential (\ref{VNO}).

The correct opposite configuration -- condensation of the Higgs field inside the core -- requires more complex scalar
potential $V$ formed from several complex fields $\Phi_i , \ i=1,2,3 $,  \cite{BurSol}.  Kinetic part of the corresponding generalized LG model differs from those of the minimal LG model (\ref{LNO}) only by summation over the fields $\Phi_i ,$
 \be {\cal L}_{GLGkin}= -\frac 14 F_\mn F^\mn - \frac 12
\sum_i(\cD_{i\m} \Phi_i)(\cD_i ^\m \Phi_i)^*  \label{GLG} ,\ee
while the potential $V$
is changed very essentially, and has to be formed by analogy with machinery of the $N=1$ supersymmetric field theory \cite{WesBag} from a superpotential function $W(\Phi_i).$\fn{It is really not only analogy, and as we shell see, only one step differs this model from the true supersymmetric Higgs model, which was obtained by Morris in \cite{Mor} with the purpose  to get supersymmetric generalization of the Witten superconducting string model \cite{Wit}.  This model was used  in \cite{BurSol,BurSol1} to describe superconducting core of the KN solution.}
  The scalar potential\fn{The signs
bar $\bar{}$ and star ${}^* $  both are used for complex conjugation.}
  \be V(r)=\sum _i F_i F_i ^*  \label{VFi} \ee
 is formed through derivatives of the function $W(\Phi_i),$
  \be F_i = \d W /\d \Phi_i \equiv \d_i W  ,\ee
where
   \be W(\Phi_i, \bar \Phi_i) = Z(\Sigma \bar
\Sigma -\eta^2) + (Z+ \m) H \bar H, \label{WLG} \ee
    $\m$ and $ \eta $ are real constants, and the special notations are introduced
   $ (H, Z, \Sigma ) \equiv (\Phi_1, \Phi_2, \Phi_3) ,$  to identify  $\Phi_1$ as the complex Higgs field
   \be H=|H|e^{i\chi} , \label{Hc} \ee
   which interacts with the KN  vector field $A_\m$ as $ \cD_{1\m} =
\nabla_{1\m} +ie A_\m .$ The fields $\Phi_2$ and  $\Phi_3$  are assumed uncharged, and
$\cD_{i\m}= \nabla_{i\m}$ for $i=2,3.$

The condition  $F_i=\d_i W =0 $
determines two vacuum states with $V=0$:

\textbf{(I)} internal vacua: $r<R-\delta $, where the Higgs field $ |H| = \eta ,$ and $ Z=-\m, \ \Sigma=0,$

and

\textbf{(E)} external vacuum state: $r>R +\delta $, where the Higgs field $ H=0,$ and $ Z=0, \ \Sigma=\eta ,$

 separated by spike of the potential $V > 0 $ in zone

 \textbf{(R)} -- a domain wall (DW), interpolating between zones \textbf{(I)} and \textbf{(E)},
 in the full correspondence with the requirements \textbf{(I),(E),(R)}.

Reduction of the corresponding LG equations to Bogomolny form is performed by minimization of the
energy density per unit area of the DW surface,

\be \m = \frac 12 \sum_{i=1}^3 ~ [ \sum_{\m=0}^3 |\cD^{(i)}_\m \Phi_i|^2 + |\d_i W|^2 ]. \label{mu} \ee

   The four dimensional DW solutions in supersymmetric LG model have paid attention in the works \cite{FMVW,HLosShifm,AbrTown,CvQRey,CvGrRey,GibTown,ChibShif}, where it  was usually  considered the static planar DWs
   positioned in (x,y) plane, with the transverse to the wall z-direction.
   However, even in the simplest case of the one field $\Phi (z)$ and one coordinate $z $,
   \be \m = \frac 12  (|\d_z \Phi|^2 + |\d_\Phi W|^2 ), \label{Phiz} \ee
   reduction of the LG equation to Bogomolny form turns out to be nontrivial, since it requires the introduction of an arbitrary phase factor $\alpha ,$ so that (\ref{Phiz}) can be equivalently presented in the form
   \be \m = \frac 12  |\d_z \Phi  - e^{i\alpha} \d_{\bar\Phi} \bar W|^2  + Re \  e^{i\alpha} \d_z W, \label{BPhiz} \ee
   which is saturated by the Bogomolny equation
   \be \d_z \Phi = e^{i\alpha} \d_{\bar\Phi}  \bar W . \label{beq} \ee

The DW forming the KN bag is much more complicated, since  first of all it is not planar, but forms
the spheroidal boundary profile of which is shown in Fig.3. Second, it is formed by three chiral
fields $\Phi_i ,$ and thirdly, the most important feature is that this DW is not static and has
non-trivial dependence on the phases of the complex fields $\Phi_i .$  The corresponding BPS
saturated solution was found in \cite{BurBag1,Bur50}, where it was shown that the phases $\alpha_i $
of the complex fields $\Phi_i $ should acquire
nontrivial dependence from time and angular coordinate \be \alpha_1 = 2\chi (t,\phi),
\quad \alpha_2 =\alpha_3 =0 , \label{alp2chi} \ee  and  the Higgs field  becomes oscillating, showing that
  just in the KN bag model the
transformation to Bogomolny form (\ref{BPhiz}) begins to operate at full power.

\begin{figure}[ht]
\centerline{\epsfig{figure=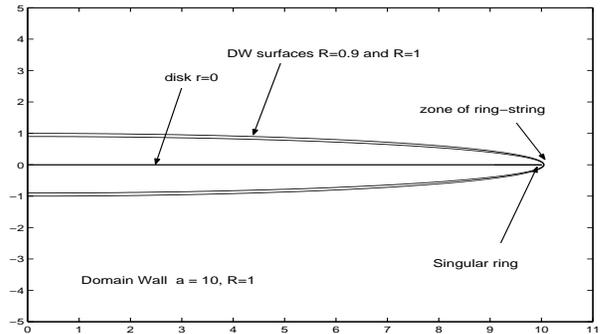,height=4.5cm,width=8cm}}
\caption{The domain wall profile (axial section) defined by the oblate spheroidal coordinate $r=R.$ }
\end{figure}

\subsection{Minimal LG model and quantization of the angular momentum}

 \noindent The non-trivial dependence (\ref{alp2chi}) is fixed in zone \textbf{(I)},
  where  the generalized LG model is reduced to minimal LG model,  and the NO Lagrangian (\ref{LNO}) leads to
  equations  \be \Box A_\m = J_\m =  e |H|^2 (\chi,_\m + e A_\m)
\label{Main} .\ee One sees that vector potential $A_\m$ acquires from the Higgs field the mass term
$m_v=e |H| ,$ and the EM field becomes short-range, with the characteristic parameter $\lambda
= 1/(e|H|) $ corresponding to the penetration depth of the EM field in superconductivity. As a
consequence, the currents vanish inside the core,  $ J_\m =0 ,$ leading to the equations \be  \Box
A_\m =0, \quad \chi,_\m + e A_\m =0 , \label{Main0} \ee showing that besides of the massive
component $A^{m_v}_\m$ which falls off receiving the mass $m_v$ from the Higgs field,  there are
also the components of different behavior.

Vector-potential of the external KN solution (\ref{Amuk}) is
\be A_\m dx^\m = - \frac {er} {r^2+a^2 \cos^2 \theta} (dr - dt - a \sin ^2 \theta d\phi ) .\label{Am} \ee
It grows near the core and takes maximal value  at the boundary of the disk, at $r = R = e^2/2m , \ \cos\theta =0 ,$
\be A^{max}_\m dx^\m = - Re \ \frac {2m} {e} (dr - dt - a d\phi ) \label{Amax}.\ee
Note, that the component $A_r$ is a perfect differential (as it is shown for example in \cite{DKS}) and can be ignored.
 At the boundary, $ A^{max}_\m$ is dragged by the light-like direction of the Kerr singular ring (see Fig.2) and the component $A^{max}_\phi $ forms the closed Wilson loop, so that \be e\oint A^{max}_\phi d\phi = 4 \pi ma  .\ee
 The right equation in (\ref{Main0}) shows that  penetrating inside the disk  vector potential determine oscillating phase of the   Higgs field as $\chi = 2m t + 2am\phi .$ The condition of multiplicity of the periods $\chi$ and $\phi $  gives $2 am = n, n=i,2,3, ..,$ which in view of  $J=ma ,$ leads to quantization of angular momentum as
 \be J =n/2,  \ n=i,2,3, ... \label{Jn}\ee
  On the other hand (\ref{Main0}) shows that phase of Higgs field
  \be H =\Phi_1  = |H|e^{i(2m t + 2am\phi)} \label{H} \ee
  oscillates with the frequency $\omega= 2m $ which
  supports extension of the components $ A^{in}_t = \frac {2m} {e}, \quad A^{in}_\phi = \frac {2ma}{e} $  inside the disk.\fn{Note, that the left massless equation (\ref{Main0}) is also satisfied, since $\Box A^{in}_t =0$ is satisfied trivially. Also, $\Box A^{in}_\phi =0 ,$ because phase $\phi$  is analytic function of  $(x+iy),$  leading to $\Box A^{in}_\phi = \d \bar \d A^{in}_\phi =0.$ These fields do not produce the field strength.} At the disk boundary (\ref{Main0}) is broken, and according (\ref{Main}) there appear
the surface currents $J_\m .$

  \begin{figure}[ht]
\centerline{\epsfig{figure=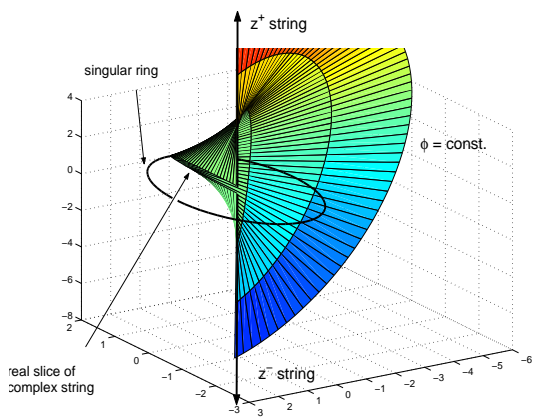,height=5.5cm,width=7cm}} \caption{\label{label} Kerr's
coordinate $\phi=const.$   Kerr singular ring drags the vector potential, forming a closed Wilson
loop along edge border of the DW.}
\end{figure}

\section{SuperBag as nonperturbative solution of the SuperQED model}

\subsection{Bosonic sector of the supersymmetric LG model}
As we noticed earlier, the generalized LG model based on the superpotential (\ref{WLG}) is not true
supersymmetric model. The difference is that the true superpotential $W$ is to be a
chiral function of the chiral superfields $\Phi_i ,$ while the scalar potential  \be V= F_i
F_i^* \ee is formed from the chiral part \be F_i^* = \d W / \d \Phi_i ,\ee but also incudes
the antichiral superpotential $W^+ (\Phi_i^+)$ depending on the antichiral superfields $\Phi_i^+ $
\be F_i = \d W^+ /\d \Phi_i^+ .\ee  These relations are retained in the bosonic sector of the
supersymmetric theory, where the fields $\Phi_i$ and $\Phi_i^+$ turn into the complex conjugate scalar
components of the superfields.

To get  full correspondence with supersymmetric theory, the fields $\Phi_i$ and $\bar \Phi_i $
in (\ref{WLG}), should be considered as independent chiral fields
$\Phi_i$ and $\tilde \Phi_i ,$ and there should also be introduced an
antichiral superpotential $W^+(\Phi_i^+, \tilde \Phi_i^+),$ which in the bosonic sector turns into
complex conjugated superpotential, built of the complex conjugated fields $\bar W(\Phi_i^*, \tilde
\Phi_i^*).$
From the complex point of view, the transition from (\ref{WLG}) to  supersymmeric
 Higgs model may be considered as \emph{complexification} of the moduli space -- analytical
extension from the real section, fixed by condition  $\bar \Phi_i = \Phi_i^* ,$ to its complex extension, the manifold with independent
coordinates $\Phi_i$ and $\tilde\Phi_i , $ supplemented with  complex conjugate coordinates  $\Phi_i^*, \ \tilde\Phi_i^* . $
Therefore, the transition to bosonic sector of the  supersymmetric generalized LG model
requires doubling of the chiral field to eliminate their degeneracy on the real slice.

Returning to the original work by Morris \cite{Mor}, where the potential (\ref{WLG}) was suggested
for super-generalization of the Witten's superconducting string model \cite{Wit}, we should double
the charged chiral fields $\Sigma$ and $\Phi ,$ and consider five chiral superfields $\Sigma_\pm,
\Phi_\pm ,$ and $Z ,$ which in Witten's interpretation of this model as the $U(I) \times U'(I)$
 Higgs field model, acquire the charges $(\pm 1, 0)$ for $\Phi_\pm ,$ and charges
for the $\Sigma_\pm $ fields as $(0, \pm 1).$ The chiral superpotential (\ref{WLG}) takes the form
\be W(\Phi_i, \tilde \Phi_i) = Z(\Sigma_+ \Sigma_- -\eta^2) + (Z+ \m) \Phi_+ \Phi_- , \label{Wchir}
\ee with identification \be  \Phi_i = (\Phi_+, \Phi_-, \Sigma_+,\Sigma_-, Z) .  \ee

The auxiliary fields
\be F_i^* = \d W /\d \Phi_i = (F_+^*, F_-^*, F_{\Sigma +}^*, F_{\Sigma -}^*, F_Z^*) \ee take the form
 \bea F_\pm^* &=& (Z + \m) \Phi_\mp , \label{Fpm} \\
  F_{\Sigma \pm}^* &=& Z  \Sigma_\mp, \label{FS} \\
  F_Z^* &=& \Sigma_+ \Sigma_- + \Phi_+ \Phi_-
-\eta^2 \label{FZ}, \eea
Vacuum expectation values of fields $\Phi_i$ for which $F_i^* =0$ give minima of the potential $V=0$
corresponding to supersymmetric vacuum states. Just as in case ({\ref{WLG}}), we obtain two isolated vacua

\textbf{(I)} $ \Phi_-\Phi_+ = \eta ^2, \quad Z= -\mu ,  \quad \Sigma_+ = \Sigma_- = 0 ;$

 \textbf{(E)} $ \Phi_- = \Phi_+ =0, \ Z= 0, \ \Sigma_+ \Sigma_-  = \eta^2 ; $

   separated by the zone

\textbf{(R)} of the positive potential
\bea \nonumber V= |\Sigma_+ \Sigma_- + \Phi_+ \Phi_- -\eta^2|^2 + |(Z + \m)\Phi_+|^2 \\
+ |(Z + \m)\Phi_-|^2 + |Z|^2 (|\Sigma_+|^2 + |\Sigma_-|^2)
 .\eea

\subsection{Transition to SuperQED model}

We note that  two oppositely charged superfields $\Phi_+ $ and $\Phi_- $ give rise to
correspondence of the supersymmetric LG model to kinetic part of the  Wess-Zumino  SuperQED model
\cite{WesBag}, \be {\cal L}_{SQEDkin} = -\frac 14 W^a W_a + \Phi_+^+ e^{eV} \Phi_+ |_{\theta \theta
\bar \theta \bar \theta} + \Phi_-^+ e^{-eV} \Phi_- |_{\theta \theta \bar \theta \bar \theta}
\label{QED kin}, \ee where $V$ is vector superfield, and $ W^a = - \frac 14 \bar D \bar D D_\alpha
V .$ In the same time, the potential part (\ref{Wchir}) corresponds to the most general
renormalizable supersymmetric Lagrangian and gives rise to nonperturbative generalization of the
SuperQED model.

 The chiral superfields $\Phi_\pm , $  are expressed  in the component form
 \be \Phi_\pm (y) = H_\pm
(y^{\m}) + \sqrt 2 \theta \psi_\pm (y^{\m}) + \theta \theta F_\pm (y^{\m}),
\label{Phi i comp} \ee as functions of the chiral coordinates  $y^\m = x^\m  + i\theta \sigma^\m \bar \theta $ and $\theta ,$
 and the scalar components $H_\pm$ are \emph{independent} Higgs fields, splitting
of the complex conjugated Higgs field of the minimal LG model in (\ref{WLG}) and (\ref{Hc}).
  Interplay of the oppositely charged Higgs fields $H_\pm $ with vector potential in zone \textbf{(I)} is defined by (\ref{Main0}) and yields
  \be H_\pm = |H_\pm| e^{\pm i\chi}, \ \bar H_\pm = |H_\pm| e^{\mp i\chi} , \  \chi = 2m t + 2am\phi ,  \label{Hpm} \ee
where the fields $\bar H_\pm$ are scalar components of the antichiral fields
\be \Phi^+_\pm (y^+) = \bar H_\pm
(y^{+\m}) + \sqrt 2 \bar\theta \bar\psi_\pm (y^{+\m}) + \bar\theta \bar\theta \bar F_\pm (y^{+\m}),
\label{Phi+ i comp} \ee
as functions of the antichiral coordinates $y^{+\m} = x^\m  - i\theta \sigma^\m \bar \theta $ and $\bar\theta .$
The corresponding nonperturbative solution with doubled Higgs fields (\ref{Hpm})
   can be obtained similar to \cite{BurBag1}.

 In the Wess-Zumino SuperQED model, the two Weyl spinors $\psi^\pm $ in (\ref{Phi i comp}) combine into
one massive Dirac spinor of the electron -- superpartner of the Higgs doublet $H_\pm ,$
\cite{WesBag}.

The nonperturbative super-bag solution generates in the core of spinning particle the flat Compton zone \textbf{(I)}, which is  free from gravity and supersymmetric, representing  the conditions for the work  of the perturbative SuperQED model, while
  the remarkable perturbative properties of the SuprQED -- "miracleous cancellations" of the component super-graphs \cite{WesBag} for a link to perturbative QED.
Note, that in the nonperturbative model of super-bag, the superpartners cannot be considered as separate
particles, and are integrated as the  superfield components of a single nonperturbative solution. The
super-bag model reveals correspondence not only with  gravity and electroweak sector of the SM, but
also with a nonperturbative version of the SuperQED model.

\section{Outlook}
We have considered principal features of the Kerr-Schild geometry which specify the supersymmetric
bag model as a new way to particle physics consistent with gravity and electroweak sector of the
SM. Two of them are principally new relative to the widespread belief:

-- the spinning KN gravity is not weak, and  becomes very strong at the Compton scale of the
particle physics,

-- compatibility between Quantum and Gravity can be achieved by means of supersymmetric
generalization of the matter sector, without modification of the Einstein-Maxwell theory.

We considered interplay of the KN gravity with  the matter sector based on the supersymmetric
generalized LG field model, which is  equivalent to supersymmetric Higgs mechanism of symmetry
breaking, and give a nonperturbative solution to generalized LG field model in the form of a
super-bag -- nonperturbative version of the SuperQED model. By conception, the 4d super-bag model
has to be soft and oscillating, similar to the conception of the superstring models
\cite{SLAC,Giles,JT}.

Due to extreme high spin/mass ratio, impact of the gravitational KN field on the structure of
space-time becomes very strong, and the consistent supersymmetric nonperturbative solutions become
very sensible to the external Einstein-Maxwell field. As a result,

a) the super-bag model creates a free from gravity Compton core of spinning particle, where the
supersymmetric vacuum state of the Higgs field provides the flat space, required for consistent
work of quantum theory;

b) the super-bag takes the shape of a strongly oblate disk forming a circular string along its
border;

c) gravitational and electromagnetic excitations of the KN solution create consistent stringy
oscillations of the super-bag in the form of traveling waves.

Many problems remain to be solved.  The closest is the so far unsolved problem of the exact nonstationary (oscillating or
accelerating) generalization of the KN solution, the problem of the consistent solutions of the
Dirac equation corresponding to confinement of quark inside the bag, and so on.

Nevertheless, the considered here features of the super-bag model are so intriguing that we risk
to state that they really give  the key to solution
of the principal problem of unification of gravity with particle physics.

 Finally, we should mention very important new aspect of this study, the direct link to the non-perturbative Wess-Zumino SuperQED model, which provides remarkable cancellations between component diagrams, presenting a link between the nonperturbative bag-like solution and the conventional technics of the perturbative QED.

\section*{Acknowledgements}
Author thanks Yu.N. Obukhov and O. Teryaev for discussion that stimulated appearance of this work, and also
 V.A. Rubakov for support on the early stage of this work. Author is also grateful to  D.V. Gal'tsov and A.A. Starobinsky for interest to this work and supporting remarks, and to K.V. Stepanyants for detailed discussion of the structure of SuperQED model.

\end{document}